\documentclass{aa}  

\usepackage{graphicx}
\usepackage{natbib}
\usepackage{txfonts}

\usepackage{hyperref}
\usepackage{amsmath}

\hypersetup{colorlinks=true,allcolors=[rgb]{0,0,0.8}}

\makeatletter
\renewcommand*\aa@pageof{, page \thepage{} of \pageref*{LastPage}}

\begin{document} 

\title{Direct detectability of tidally heated exomoons \\ by photometric orbital modulation}

   \author{E. Kleisioti \inst{1}\fnmsep
         \inst{2}
         \and 
         D. Dirkx \inst{2}
         \and 
         X. Tan \inst{3}\fnmsep\inst{4}
         \and
         M. A. Kenworthy\inst{1}}

   \institute{Leiden Observatory, Leiden University, P.O. Box 9513, 2300 RA Leiden, The Netherlands\\
              \email{kleisioti@mail.strw.leidenuniv.nl}
         \and
             Faculty of Aerospace Engineering, TU Delft, Building 62 Kluyverweg 1, 2629 HS Delft, the Netherlands
         \and
            Tsung-Dao Lee Institute, Shanghai Jiao Tong University, 520 Shengrong Road, Shanghai, 200127, People's Republic of China
        \and
        School of Physics and Astronomy, Shanghai Jiao Tong University, 800 Dongchuan Road, Shanghai,200240, People's Republic of China}

   \date{Received \today; accepted XX}

 
  \abstract
   {}
   {We investigate whether volcanic exomoons can be detected in thermal wavelength light curves due to their phase variability along their orbit.
   The method we use is based on the photometric signal  variability that volcanic features or hotspots would cause in infrared (IR) wavelengths, when they are inhomogeneously distributed on the surface of a tidally heated exomoon (THEM).}
 {We simulated satellites of various sizes around an isolated planet and modeled the system's variability in two IR wavelengths, taking into account photon shot noise.
   The moon's periodic signal as it orbits the planet introduces a peak in the frequency space of the system's time-variable flux.
   We investigated the THEM and system properties that would make a moon stand out in the frequency space of its host's variable flux.}
   {The moon's signal can produce a prominent feature in its host's flux periodogram at shorter IR wavelengths for hotspots with temperatures similar to the ones seen on the Jovian moon, Io, while the same moon would not be identifiable in longer IR wavelengths.
   By comparing observations at two different wavelengths, we are able to disentangle an exomoon's signal from the planet's one in the frequency domain for system distances up to $\sim$10 pc for Mars-sized exomoons and even further for Earth-sized ones for transiting and non-transiting orbital inclinations.}
   {This method enlarges the parameter space of detectable exomoons around isolated planetary mass objects and directly imaged exoplanets, as it is sensitive to Io-Earth sized exomoons with hot volcanic features for a wide range of non-transiting orbital inclinations.
   Exomoon transits and the detection of outgassed volcanic molecules can subsequently confirm a putative detection.}

\titlerunning{Direct detectability of tidally heated exomoons by photometric orbital modulation}

\authorrunning{Kleisioti et al.}


   \maketitle
%

\section{Introduction}

Solar System planets host a large number of moons, with great diversity; from the gas giants' icy satellites, to the volcanic moon, Io, and the only moon with a substantial atmosphere, Titan.
In the Solar System, moons present a wide variety of geological and orbital properties; however, their detection around exoplanets still remains to be confirmed.
 Given the large number of discovered exoplanets, their detection is ever more likely. 

Searches for extrasolar satellites are unveiling upper limits on exomoon properties from a non-detection \citep{Vanderburg_2021, Ruffio_2023} or effects that are associated with the existence of a moon, such as the detection of volcanic species and gaps in ring systems \citep{Oza_2019, Kenworthy_2015}. Candidate exomoons have been proposed \citep{Teachey_kipping_2018, Fox2020, Kipping_2022}; however, their detection has been put into question \citep{Rodenbeck2018, Heller2019, Kipping_2020, Heller_2023}.

Several exomoon detection methods around close-in planets have been suggested, following the success of the exoplanet transit method.
Transit timing \citep{Sartoretti_schneider_1999, simon2007} and transit duration variations \citep{kipping2009} are based on the effects that an exomoon has on planetary transits due to gravitational interactions.
Furthermore, \cite{heller2014} suggested detection via statistical analysis of the stellar photometric  transit signal.
Exomoon atmospheres can also be detected via transit spectroscopy \citep{Kaltenegger_2010}. 
However, these methods are only applicable to close-in planets, which according to orbital evolution models are less likely to host satellites \citep{Trani_2020, Dobos_2021}.

Directly imaged (DI) exoplanets, on the other hand, are further away from their parent star due to the challenges brought by star-light suppression at small angular distances.
The formation pathways of DI exoplanets are still unclear \citep[e.g., ][]{bohn_2021}, and could be better constrained with a satellite detection around them.
Exomoon detection methods around DI exoplanets include direct imaging of an unresolved satellite, planet radial velocity measurements, planet astrometry, and transits (\cite{cabrera_schneider_2007, Agol_2015, Vanderburg_2018, vanWoerkom2024}).
With current instrumentation, these methods are limited to detecting binary-like satellites \citep{lazzoni2022}. 
Such techniques can also be applied to brown dwarfs (BDs) or isolated planetary-mass objects (IPMOs) \citep{Limbach_2021}, since high-contrast imaging is not necessary to subtract the stellar light. 

Direct detection via thermal excess of tidally heated exomoons (THEMs), heated by tidal interactions along an eccentric, short-period orbit has also been proposed \citep{Limbach_2013}.
The latter requires tidal heating to take place in the interior of the moon.
Tidal heating is present in Solar System bodies, the most prominent example being that of Io  \citep[\textit{e.g., }][]{Moore_2003,Behounkova_2023, Rovira_2022}.
Most studies on the direct detection of THEMs have considered a uniform temperature distribution over the moon's surface to evaluate the temperatures that can be reached by tidal interactions and assess potential detections of thermal excess \citep{Dobos_2015,Rovirra_2021,  Kleisioti_2023}. 
However, the latter assumption does not take into account the temperature inhomogeneity on the surface of our only known example of a THEM, Io \citep{ DEPATER2017, de_Kleer_2019}. 
For example, Loki Patera, one of the largest and most persistent hotspots on Io's surface, with a surface area equal to 21500 km$^{2}$, or  0.05 \% of its surface \citep{DAVIES201567}, is responsible for 10 percent of Io’s heat flow \citep{veeder1994}.

 \cite{1jager2021} suggest that the temperature distribution in a THEM will be uneven.
 The authors claim that this uneven temperature distribution can enhance the direct detectability of exomoons. 
Motivated by the latter, we argue that such a hotspot distribution would produce a periodic signal in thermal infrared (IR) light curves of BDs, DI exoplanets, and IPMOs, and we explore the conditions under which this periodicity is detectable. 
The detection of extrasolar bodies via thermal periodicity has previously been suggested for the detection of rocky asteroids \citep{Lin_2014} and exoplanets around white dwarfs (WDs)  \citep{Limbach_2022}. 
The authors considered detection by measuring amplitude variations in WD light curves from rocky bodies' phase curves due to the thermal emission differences between the dayside and the nightside.
In addition, \citet{Forgan2017} suggested that moons could be detected via periodicity detection in exoplanetary transit phase curves.
Thermal phase curves have been used for the characterization of rocky exoplanets around main-sequence stars \citep{Demory_2016, Zieba_2022, 2023jwst.prop.3077G}.
To quantify the feasibility of our proposed method, we simulated phase curves of THEMs with hotspots, adding realistic host variable signals.
We show that the thermal output of hotspots can produce detectable signals with MIRI in their hosts' IR light curves by  extracting the signal of the moon from the host's periodogram.  

This work is organized as follows: in Section \ref{photometric_signal_variability}, we present our models for exomoon and host signal variability. Section \ref{Detectability} discusses our method of extracting the exomoon signal from the host's periodogram. In Section \ref{results}, we present our results, firstly for a system at 3 pc and later on for THEMs with different properties at several distances. Finally, in Section \ref{discussion}, we discuss possible synergies of our method with other exomoon detection methods and ways to confirm a detection.

\section{Photometric signal variability} \label{photometric_signal_variability}

In this section we present the models that we use for the photometric variability over time of a volcanic moon and its host.
We then compute the combined light curve of the system by adding the two light curves.

\subsection{Exo-Io modeled photometric light curves}

Volcanic activity on Io is the result of tidally induced heating in the interior of the moon.
Over the years, Io's volcanic activity has been monitored \citep[\textit{e.g.,} ][]{DEPATER2017,CANTRALL2018,de_Kleer_2019, tate2023spatiotemporal} and the spatial distribution of volcanoes \citep{bartolić2021occultation} has been used to infer tidal heating parameters in the interior \citep{Davies2024}. 
Measured hotspot temperatures on Io reach up to $\approx$ 2000 K \citep{mcEwen1998}.
A larger concentration of  hotspots was found by \cite{Zambon2022} in  the  polar regions of Io; however, recently \cite{Davies2024} concluded that Io’s polar volcanoes are about the same in number per unit area but are less energetic than those at lower latitudes.

\cite{DEKLEER2016} divided Io's hotspots into two main categories depending on their temporal behavior and temperature. 
``Sudden brightening events'' are characterized by a sudden high thermal output, which steadily decays over a timescale of $\approx$1  month.
On the other hand, ``persistent hotspots'' are active for more than $\approx$ 1 year and show low temporal variability with typically lower thermal outputs.
Loki Patera, one of the most persistent hotspots on Io's surface \citep{de_Kleer_2019}, presents characteristics of both categories, showing both stability over time and eruption events, with temperatures of up to $\approx$ 540 K \citep{DEPATER2017}.

\begin{figure}[!htb]
    \centering
    \includegraphics[width=0.98\columnwidth]{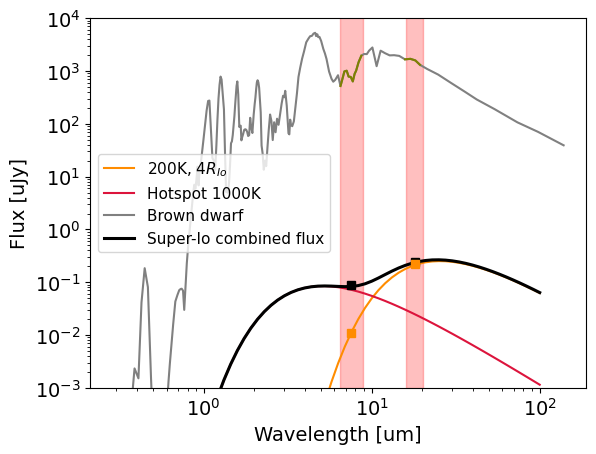}
    \caption{Brown dwarf spectrum with $T_{eff}=$ 600 K and a THEM's spectrum with $T_{m}=$ 200K, $T_{h}=$ 1000K, and $R_{m}= R_{Earth}$ at 3 pc.
    The two red rectangles show the wavelength ranges of MIRI filters F770W and F1800W.
    The black and orange points correspond to the integrated flux coming from a moon when the volcano is in and out of our line of sight, respectively.
    The difference between the two configurations is larger in filter F770W, as is shown by the orange and black points.}
    \label{fig:moonandplanetfluxl}
\end{figure}

We modeled THEM light curves, assuming moons that are tidally locked to their host, with different radii, $R_{m}$, varying from Io-sized bodies to Earth-sized ones, with a surface temperature baseline of $T_{m}$. 
We generated a spherical harmonics map with $starry$ \citep{starryLuger} to model the nonhomogenous surface of THEMs. 
We added the expansion of one and three top hat function(s) on the map to simulate one or three circular spots, respectively, with radius $R_{h}$ and temperature $T_{h}$. 
We chose a spherical harmonics degree $= 30$, as we modeled small features on the moon's surface, due to the small $R_{h}/R_{m}$ ratio. 

\begin{figure}[!htb]
    \centering
    \includegraphics[width=1\columnwidth]{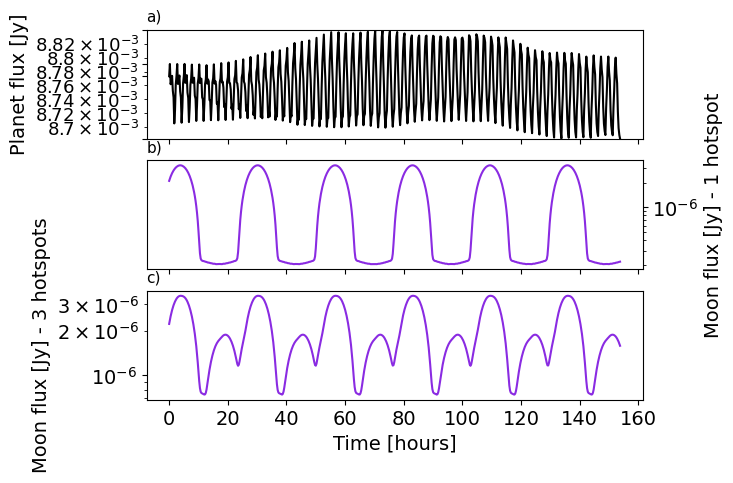}
    \caption{Modeled F770W flux variabilities of (a) the host and an Earth-sized, $T_{m}=$200 K, $T_{h}=$1000K moon with (b) one or (c) three hotspots at 90 degrees inclination at 3 pc. The overall observation time (including both filters) assumed in this work is 140h.}
    \label{fig:moonandplanetvariabilityl}
\end{figure}

Figure \ref{fig:moonandplanetfluxl} shows the spectral energy distribution of a moon with $T_{m}=$ 200K and $T_{h}= $1000K. 
The resulting flux variability of a rotating moon would differ from longer wavelength filters to shorter wavelength ones; the flux difference induced by the visibility of the volcano  is larger in the JWST/MIRI filter F770W than in F1800W, as is shown by the black and orange points, respectively.
Panels b and c in Figure \ref{fig:moonandplanetvariabilityl} show the moon's temporal light curve with Io's orbital period and in an edge-on orbit, at a distance of 3 pc. When modeling moons larger in size than Io, we kept the ratio, $\frac{R_{h}}{R_{m}} = \frac{R_{Loki Patera}}{R_{Io}} = 4.54 \cdot 10^{-2} $, constant, assuming Loki Patera's  surface area to be equal to $21500 km^2$, as it was measured to be by \citep{DAVIES201567}.

\subsection{Host modeled photometric light curves} \label{host model}

Exoplanets and BDs are expected to present variability due to atmospheric inhomogenities (e.g., clouds) combined with their rotational modulation. 
Several surveys have demonstrated that this variability is common among BDs \citep{Metchev_2015, Vos_2022} and is often well described by a semi-periodic signal or a combination of several sine functions for observations in one epoch \citep{biller2017, apai2017, Artigau2018}. 
Some observed light curves also evolve in time \citep{Artigau_2009, Yang_2016, Karalidi_2016}. 
Recent surveys of young, low-mass BDs measure a variability between 2 and $\approx$ 35 hours with a median of 12 hours \citep{Vos_2022}.
Simulated normalized light curves of such bodies have amplitudes from 0.5 percent to a few percent, which is consistent with observations \citep{Radigan_2014, vos_2019}. 
Using atmospheric general circulation models (GCMs) of BDs, \citet{Tan_2021} found that faster rotators typically present smaller variability amplitudes.
Finally, lower-mass BDs are likely more variable than their more massive counterparts \citep{vos_2019}.
BDs and DI exoplanets share color and magnitude similarities \citep{faherty_2016} and from the perspective of planetary atmospheric dynamics they are in the same self-luminous dynamical regime \citep{Showman_2020}.

Most IR light curve datasets of BDs or IMPOs are measured continuously for short timescales (e.g., 5-20 hours \cite{vos_2019, esplin2016}) or with lower cadences (e.g., days or weeks \citep{Brooks_2023}). 
Thus, any long-term evolution is not accurately measured in such systems, with the exception of \cite{Apai_2021}. 
Longer datasets are needed for the detection of moons, which can have orbital periods much larger than available datasets. 
In this work, we use GCMs to simulate IPMOs' bolometric light curves and model their long-term evolution. The IPMO circulation model in this work has been updated from that in \cite{Tan_2021} and \cite{Tan_2021b} by using a more realistic non-gray radiative transfer and assuming equilibrium chemistry and a solar atmospheric composition (Tan et al. in prep). 
We also computed spectroscopic light curves to estimate the variability at different MIRI filters by post-processing the 3D temperature and cloud IMPO circulation model fields with PICASO \citep[][ see Appendix A]{Batalha_2019,ADAMS2022, Robbins_Blanch_2022}. The IMPO circulation model and PICASO have the same opacity source and method of computing radiative transfer, which ensures the accurate spectroscopic post-processing of the circulation.
 
For our spectroscopic light curves, we substituted the amplitude of the calculated bolometric flux with the characteristic amplitude from the spectroscopic light curve within a given bandpass.
This assumes that the light curve periodicity at wavelengths between 6-20$\mu$m is similar to that of the bolometric light curve, a behavior that has been observed in measured spectroscopic light curves \citep{apai2017,esplin2016,  Yang_2016}.

 Figure \ref{fig:moonandplanetvariabilityl}a shows the calculated F770W/MIRI variability of a 2.5 h variable host, with an effective temperature of $T_{eff}=$600 K and a surface gravity of $g=$100 m/s$^2$.
Our goal is to demonstrate that we can distinguish the signal of a volcanic companion in a realistic IPMO light curve. 
When simulating light curves of systems viewed near edge-on, we take into account only the intrinsic variability of the IPMO and companion signals, without considering any transits or secondary eclipses, as
we expect them to enhance the periodicity of the signal, and thus the detectability compared to a non-transiting moon.
We present results of simulations using the modeled IMPO as a host and argue that the same method can be used to find companions around lower-mass IMPOs and DI planets.
When referring to the companion of DI exoplanets and IPMOs, the term “exomoon” can be used; however, it should be noted that there is a blurred line between exoplanets, IPMOs, and BDs \citep{teachey2024detecting}. When this method is used to find companions around isolated and more massive BDs, the term “volcanic companion” is more accurate. 

\section{Detectable moon signals in hosts' periodograms} \label{Detectability}

To assess whether an exomoon would produce a detectable signal peak when taking time-series observations of a host, we firstly added the time-variable flux of moons with different $T_{h}, R_{h}, R_{m}$, and $T_{m}$ to the flux of the 2.5 h variable host in two MIRI filters.
It should be noted that for a $T_{h}$ range between 300 and 2000K the optimal filter choice would change, since the blackbody peak changes. 
We then integrated over 0.5h observations, alternating between each filter after every integration, and added shot noise to the combined system flux. 

\begin{figure}[!htb]
     \centering
     \includegraphics[width=0.98\columnwidth]
     {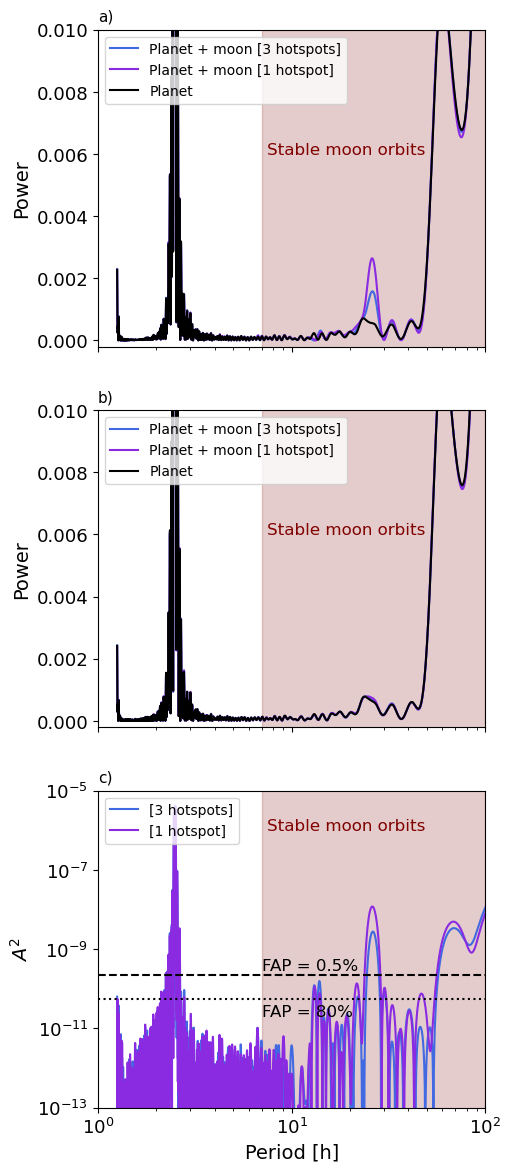}
     \caption{(a) Lomb-scargle periodogram for MIRI filter F770W. (b) Lomb-scargle periodogram for MIRI filter F1800W. The THEM signal is diluted in longer thermal wavelengths. (c) A$^{2}$ as a function of the moon's orbital period for the 2.5 h variable IPMO and an Earth-sized exomoon with $T_h=$ 1000 K at an edge-on orbit at 3 pc. The dashed and dotted black lines denote $FAP = 0.5\%$ and  $FAP = 80\%$. The red area shows the stable moon orbit region, between the host's Roche limit and the Hill sphere.}
     \label{fig:perioddogram}
 \end{figure}

Panels a and b of Figure \ref{fig:perioddogram}  show the Lomb-scargle periodogram of the combined fluxes in filters F770W and F1800W, respectively.
 The black curves demonstrate the periodogram of the host's light curve, whereas the purple (one hotspot) and blue (three hotspots) are the result of the combined system flux with an Earth-sized moon that has $T_{h}=$ 1000K at a 26.4h orbital period.
 There are several peaks on the planet's periodogram, with the most prominent one being equal to the planets intrisic variability at 2.5 h.
 The planet's periodogram (black) also exhibits several smaller peaks at larger periods that can hinder the moon signal \citep{Apai_2021}.
 This is expected to affect the detectability of moons with different orbital periods, which we explore in Section \ref{RESULTS1}.

In the combined periodogram (purple) in filter F770W there is a clear peak at $\approx 26.4 $ hours that is not present in the host's periodogram (black).
 The same moon produces a much smaller signal in the F1800W filter, since it does not capture the high temperature profile of the hotspot, as is seen in Figure \ref{fig:moonandplanetfluxl}. 
 The latter discrepancy between the two filters enables the moon's signal to be disentangled from the planet's one.

To better assess the differences between the two filters and the detectability of moons, we also defined a quantity, $A$, as follows:

\begin{equation} \label{differnece_filters}
    A = P_{F770W} - P_{F1800W}
,\end{equation}
where $P$ is the periodogram's power for the two different filters.
Figure \ref{fig:perioddogram}c shows $A^2$ as a function of the moon's orbital period, where the peak due to the input orbital period (26.4h) is the most prominent feature. The vertical black lines correspond to false alarm probabilities (FAPs), calculated according to the null hypothesis of non-varying data with Gaussian noise. 

\begin{figure}[!htb]
     \centering
     \includegraphics[width=0.98\columnwidth]{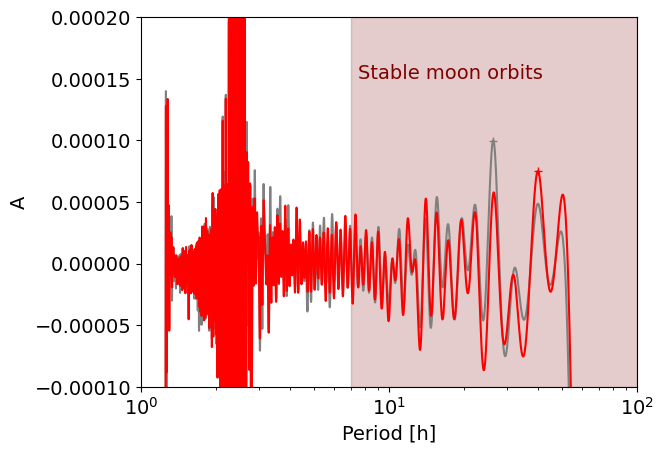}
     \caption{Two instances of $A$, which is subject to shot noise, for the 2.5 h variable planet and a THEM with $T_{h}=   $600 K, at 3 pc, and 90 degrees inclination.
     The instance shown with a gray line correctly identifies the input moon orbital period (26.4 h), in contrast to the instance represented by a red line.The red area shows the stable moon orbit region, between the host's Roche limit and the Hill sphere.}
     \label{fig:shotnoise}
 \end{figure}

Since photon noise is a stochastic process, Figure \ref{fig:perioddogram} shows only one of the possible instances that we might expect from such a system.
Figure \ref{fig:shotnoise} shows the same system with $T_{h}=$ 600K at two different instances of shot noise, one where the moon's peak is the most prominent one (gray) and one where the shot noise interferes with the signal, resulting in a spurious feature (red).
To assess the effect of the photon noise on the observability of the exomoon signal, we performed the same calculation $10^3$ times (see Appendix B) and defined a true positive rate as the ratio of the instances in which the most prominent peak is caused by the moon over the total number of instances.
The relationship between the FAP and the true positive rate is shown in Appendix C.

\begin{figure}[!htb]
    \centering
    \includegraphics[width=0.98\columnwidth]{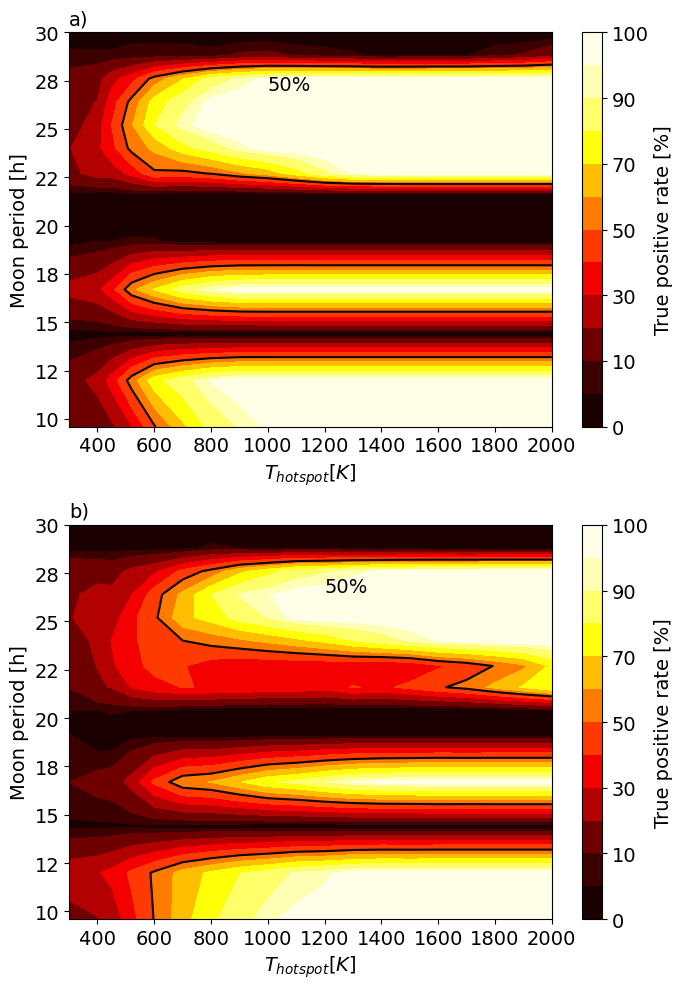}
    \caption{True-positive ratio for an Earth sized moon around the 2.5 h variable planet as a function of the moon's orbital period and $T_{h}$.
    The system is assumed to be at 3 pc for an (a) edge-on and (b) 30$^{\circ}$ inclined orbit.}
    \label{fig:9030deg}
\end{figure}

The signals of THEMs would appear between periods that can theoretically be constrained by taking into account their orbital stability.
%
In a THEM search via photometric orbital modulation, moon peaks would appear in periods larger than the Roche limit of a particular system.
In the following simulations, we took the smaller period limit at which a moon's signal can appear in the planet's periodogram, which is equal to the Roche limit for the modeled IPMO with radius $R = 1.006 R_J$  and mass $M = 4 M_J$.

\section{Results} \label{results}
In this section we present our results, firstly for the modeled 2.5 h variable planet at 3 pc presented in Section \ref{host model}.
We then expand on the potential detectability of THEMs for some typical systems.

\subsection{Tidally heated exomoons around a 2.5 h variable planet at 3 parsecs} \label{RESULTS1}
Figure \ref{fig:9030deg}a shows the true positive ratio of an Earth-sized moon as a function of its orbital period and hotspot temperature ($T_{h}$) at a 90$^{\circ}$ inclination.
A black line is overplotted for a true positive ratio = 50 \%.
For the majority of orbital periods, an Earth-sized moon's signal has a true positive rate of $50 \%$ for $T_{h}=$ 550K. 
We note that F770W is not the optimal filter for the detection of bodies with $T_{h}$ in the entire shown parameter space; however, such an analysis is out of the scope of this work.

There are some dark regions in Figure \ref{fig:9030deg} that represent a true positive rate equal to zero.
These correspond to periods where the intrinsic variability of the host interferes with the moon's signal.
Moons with orbital periods within this range would not produce a detectable peak in the host's periodogram regardless of their hotspot properties.
The periods at which this occurs cannot be predicted a priori; however, observations at a longer IR wavelength (F1800W) can constrain them, assuming that there is a similarity between the host's variability periods in the two bands (Appendix A).

As is seen in Figure \ref{fig:9030deg}b, a smaller inclination ($30^{\circ}$) does not significantly affect the observable range of moon orbital periods.
However, the hotspot temperatures that produce detectable moon peaks are larger: $\approx$ 600 K for an Earth-sized moon with $T_{m}=$ 200 K.

\subsection{Applicability to systems at larger distances}
\begin{figure}[!htb]
    \includegraphics[width=0.98\columnwidth]{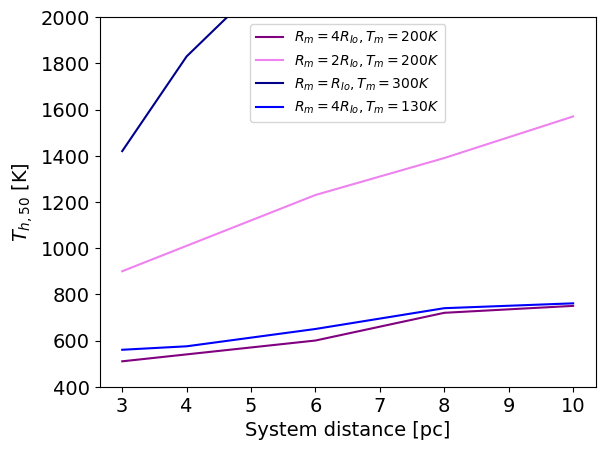}
    \caption{$T_{h,50}$ as a function of the system's distance for the modeled 2.5 h variable planet with a 600 K effective temperature and different moon properties (orbital period $=$ 27 h). For moon sizes between 2-4 $R_{Io}$, a moon can be detected even for moderate $T_{h}$ and $T_{m}$ in 140 h observations. For very nearby systems, an exo-Io ($R_{m} = R_{Io}$) can be detected for high $T_{h}$ ($>$1400 K) }
    \label{Th_50vsdistance}
\end{figure}

To evaluate the limits of this method in terms of system distance, as well as the moon and hotspot parameter spaces that lead to detectable signatures, we calculated the $T_{h}$ that results in a true positive rate $=50\%$ ($T_{h,50}$) for a range of $R_{m}, T_{m}, R_{h}$ (Figure \ref{Th_50vsdistance}).
For a system distance of $<$5 pc, exo-Ios as small as $R_{m} = R_{Io}$ can be detected for high hotspot temperatures ($>$1400K) in 140 h observations. 
For larger moons between 2 and 4 $R_{Io}$, a signal can be detected even for moderate $T_{h}$ ($\approx$ 900-1500 and 500-700K, respectively).  

\section{Discussion} \label{discussion}

A THEM with hotspot temperatures similar to the ones seen on Io can produce a detectable signature in JWST time-series data of IPMOs.
We discuss whether a detection can be justified from such a signal in putative exomoon hosts' periodograms, which other sources can produce such signatures, and factors that need to be taken into account when extending the results in different systems.

\subsection{Other sources of such signals and synergies with other exomoon detection methods}

This paper focuses on the possibility of THEM signatures being present in MIRI/JWST light curves.
Assessing the probability that a particular peak within the stable moon orbital period range is from a putative exomoon is out of the scope of this work.
The latter would depend, among other confounding effects, on the planet's intrinsic variability itself, which cannot be assumed a priori.
However, there are strategies that can be used to follow up on candidate exomoon signals.

Firstly, additional observations over long periods or within a period of time would allow for the planet's variability to change over time.
In contrast, the exomoon signal would always appear at the same period, its orbital period, even if the location and  intensity of the hotspots change.
A factor that might affect the interpretation of a signal is the integration time of each observation or the sampling frequency, as it can introduce false-positive signals.
Repeating an observation with a different sampling frequency can bypass these false positives, and allow the exomoon signal to persist.

Other exomoon detection methods can serve as confirmation of the existence of a THEM or vice versa. 
If a THEM is large enough and the orbital inclination is within the range of transiting inclinations, it can produce detectable transits in the photometric light curves of BDs or IPMOs \citep{Limbach_2022}.
However, the presented hotspot variability method is sensitive to exomoon detections in a much wider range of inclinations compared to the exomoon transit method.
Detection via IR excess \citep{Limbach_2013, Kleisioti_2023}, the detection of alkalines \citep{Oza_2019}, or radio signals induced by planet-moon interactions \citep{Narang_2023} can also contribute to constraining the existence of a moon around a BD or DI exoplanet.
Finally, the method of spectroastrometry can also contribute to the confirmation of a THEM detection, as it will be sensitive to similar, temperately warm moons with the upcoming Extremely Large Telescope (ELT) \citep{vanWoerkom2024}.
Radial velocity measurements of the planet with current instrumentation are sensitive to binary-like moons \citep{lazzoni2022} or to mass ratios much larger than the ones seen in the Solar System gas giants and their satellites \citep{Ruffio_2023}.

\subsection{Hotspot time variability and spatial distribution on Io and exo-Ios}
This method is subject to the frequency of eruptions on a putative THEM. 
Persistent volcanoes, like those seen on Io, are more likely to be present on a THEM's surface, since they are stable for more than $\approx$ 1 year; however, they have lower thermal fluxes. 
``Sudden brightening events,'' with higher temperatures, fade over a timescale of $\approx$ 1 month, which is much smaller than the observation time assumed in this work (140h), and thus also likely to be detectable.

The spatial distribution of volcanoes needs to allow enough variability in a moon's phase curve and can boost or hinder a moon's signal. 
A homogeneous distribution along the surface of hotspots with similar properties would produce a flat moon phase curve.
On the other hand, two hotspots on the same hemisphere (even if they are smaller than the modeled sizes) would enhance the moon's periodicity amplitude.

\subsection{Possibility of hotspot detection around cooler planets}

Our current models assume a host planet temperature of 600 K.
We note, however, that if the host planet is colder than this, then cooler and smaller moons will be detectable too, since the fraction, $\frac{F_m}{F_p}$, will be similar for colder volcanoes.
Ultimately, the increased fractional photon shot noise from the smaller fluxes from the host and exomoon will require longer integrations to detect the moon at a similar significance.

\begin{figure}[!htb]
    \centering
    \includegraphics[width=0.98\columnwidth]{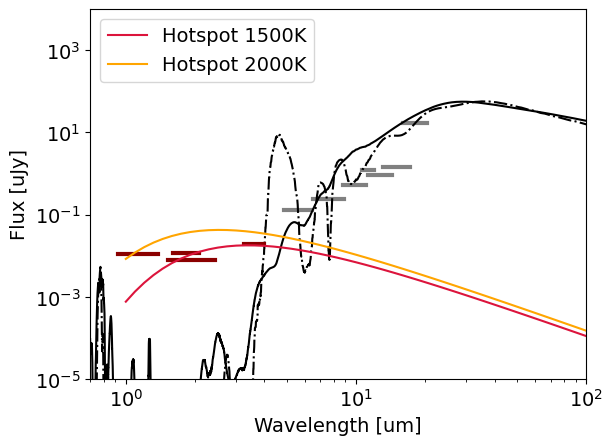}
    \caption{
    Hotspots with radii of $R_{h}$ = $R_{Loki Patera}$ at 3 pc, around a cold undetectable host of 150 K with (dashed line) or without (continuous line) clouds \citep{Morley_2014}. The vertical red and black lines correspond to the 10$\sigma$, 10000 s detection limits of NIRCam \citep{Rieke_2023} and MIRI \citep{Glasse2015}, respectively.}
    \label{fig:ppm}
\end{figure}

In the case of host exoplanets with low effective temperatures, the possibility exists that the planet itself is not detectable at all, while the hotspots on the warm end of the hotspot parameter space can exceed NIRCam's sensitivity at small distances from the Earth (see Figure \ref{fig:ppm}b).

\section{Conclusions}
We modeled the IR variability of a 600 K isolated planet and added phase curves of THEMs with different hotspot, physical, and orbital properties.
We conclude that a THEM with hotspot temperatures similar to the ones seen on the Jovian moon, Io, can produce detectable signatures in the host’s IR periodograms for a wide range of orbital inclinations.
For some exomoon orbital periods, which would depend on the planet's intristic variablity and cannot be known a priori, a detection is not possible independently of the hotspot properties.
Synergies with other methods, such as exomoon transits and spectral signatures due to volcanism, can contribute to confirming the presence of an exomoon.
When taking into account shot noise, this method can detect Io-sized satellites with volcanic temperatures $>$1400 K for distances $<$5 pc, and Mars to Earth-sized moons further away for moderate volcanic temperatures around a 600 K planet.
The latter detection limits can be looser for moons around colder hosts. 
If a host is cold enough, direct detection of a volcano is possible with NIRCam at 3 pc.

\begin{acknowledgements}
The authors thank the anonymous referee for their comments that improved the quality of the manuscript and C. Morley for providing the 150K planetary spectra.This research has been supported by the PEPSci Programme (Planetary and ExoPlanetary Science Programme), NWO, the Netherlands.
\end{acknowledgements}


%
%

\bibliography{source}
\newpage

\begin{appendix}
\onecolumn

\section{Planet spectral variability}
\begin{figure}[!htb]
    \centering
   \includegraphics[width=0.8\columnwidth]{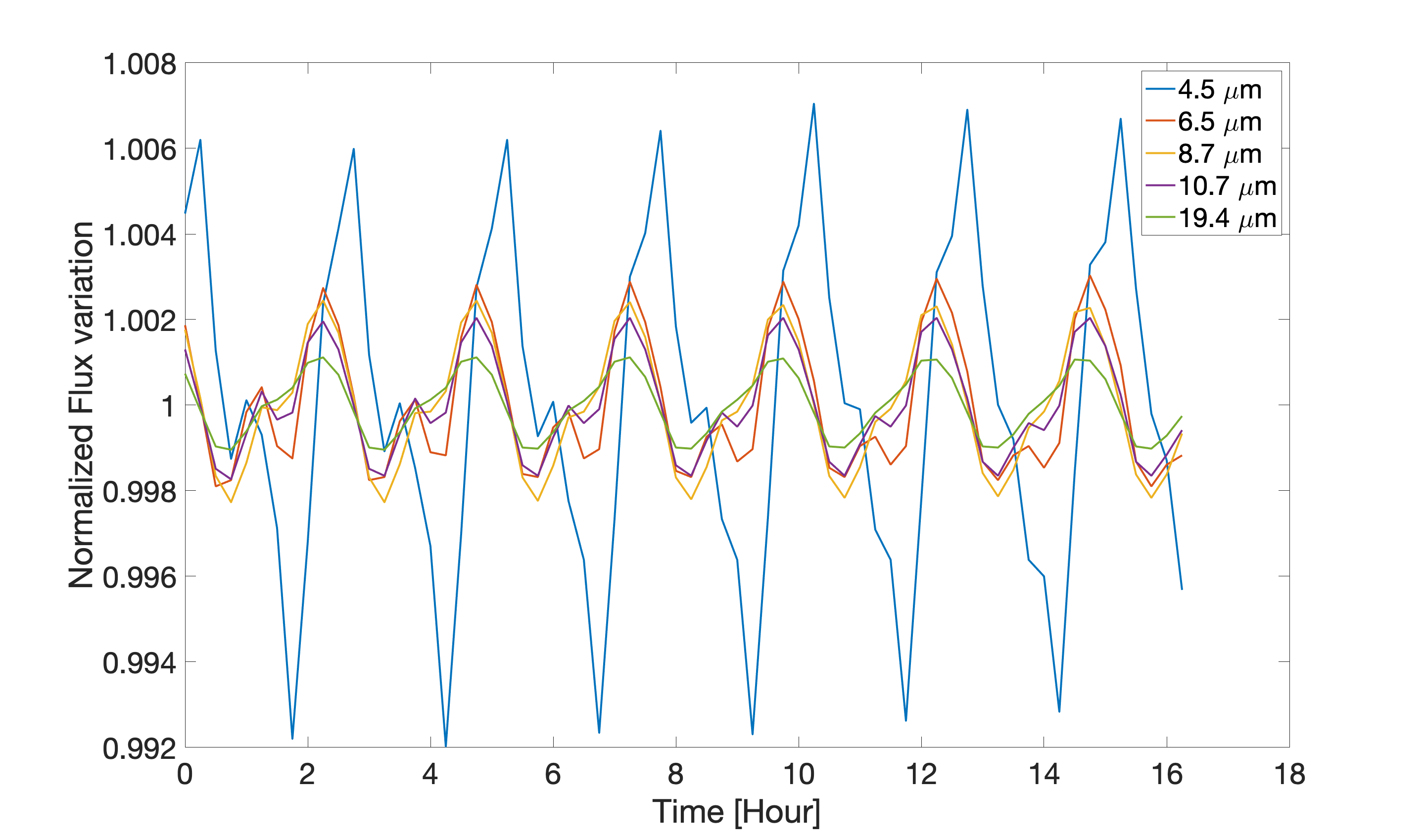}
    \caption{Host spectroscopic variability for an edge-on view. The period of variability remains the same, while the amplitude of variability changes as a function of wavelength.}
    \label{fig:my_label1}
\end{figure}

\begin{figure}[!htb]
    \centering
   \includegraphics[width=0.8\columnwidth]{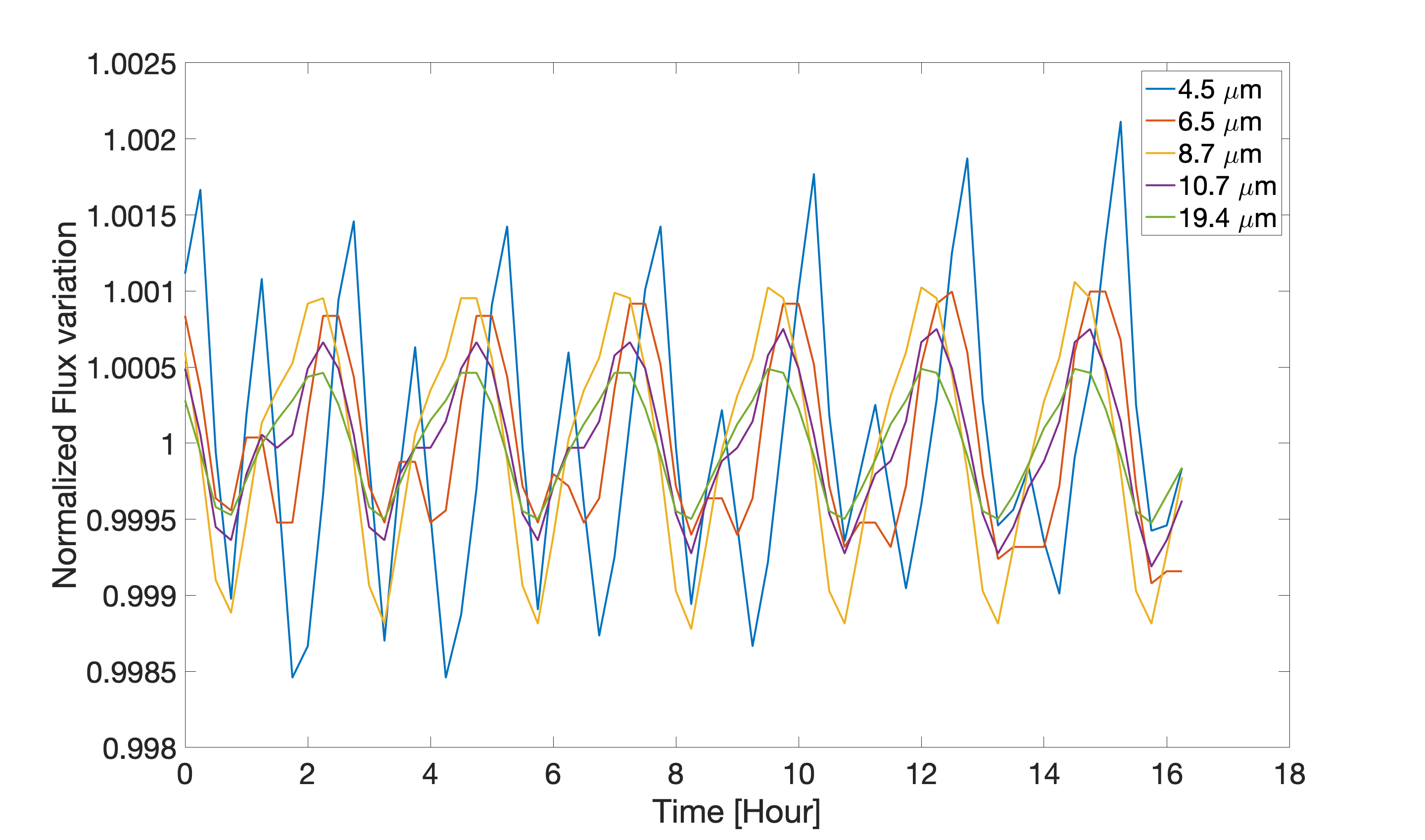}
    \caption{Host spectroscopic variability (30$^{\circ} inclined$).
    The period of variability remains the same, while the amplitude of variability changes as a function of wavelength.}
    \label{fig:my_label2}
\end{figure}

Figures \ref{fig:my_label1} and \ref{fig:my_label2} show the normalized flux variations with time, calculated by the IPMO circulation model for an edge-on and a $30\deg$ inclined orbit.

\section{True positive rate as a function of number of instances}
\begin{figure}[h!]
    \centering
   \includegraphics[width=0.6\columnwidth]{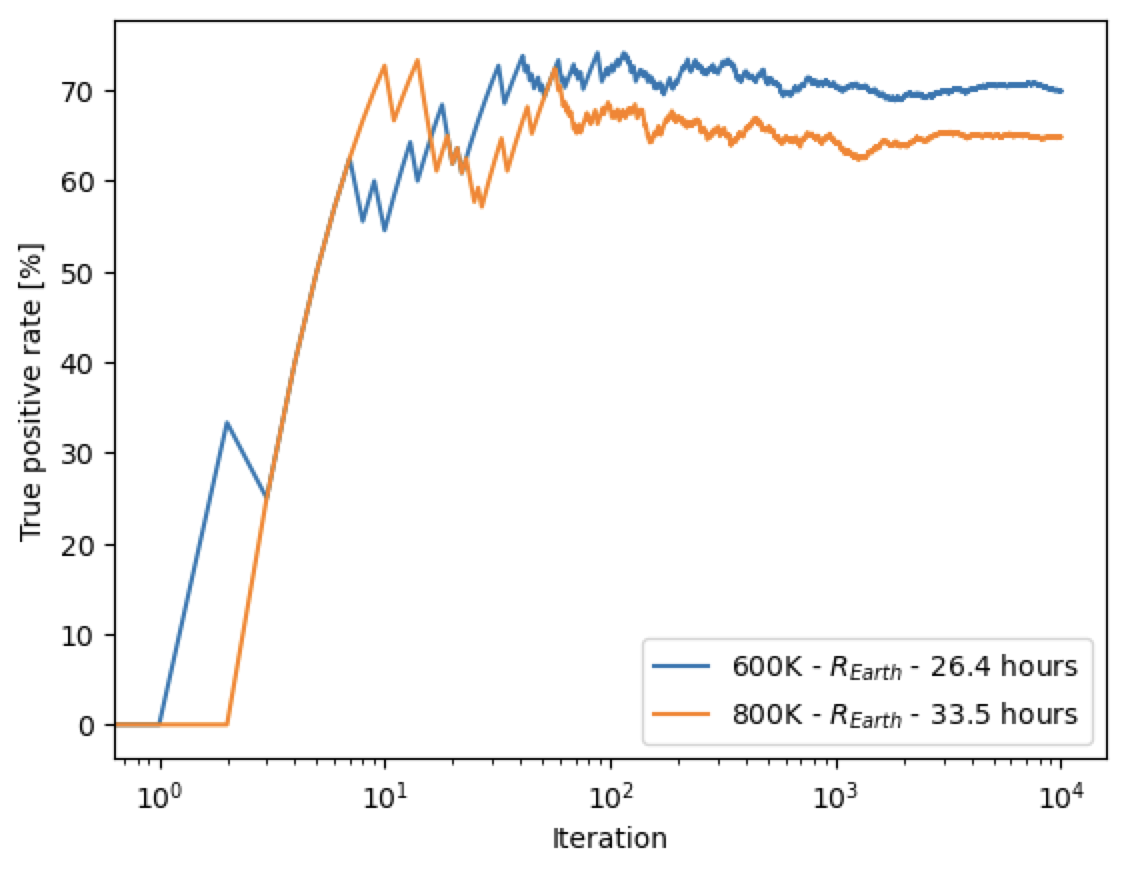}
    \caption{True positive rate as a function of the number of shot noise instances for an Earth-sized moon with $T_{h}=$ 600 K, an orbital period of 26.4 h (blue line) and $T_{h}=$ 800 K, and an orbital period of 33.5 h (orange line). }
    \label{fig:my_label}
\end{figure}

Figure \ref{fig:my_label} shows the true positive rate as a function of the number of shot noise instances (iterations). We chose $10^3$ iterations as the number of iterations for the simulations shown in Figures \ref{fig:9030deg} and \ref{Th_50vsdistance}.

\section{False alarm probabilities as a function of the true positive rate} 

\begin{figure}[h!]
    \centering
    \includegraphics[width=0.5\columnwidth]{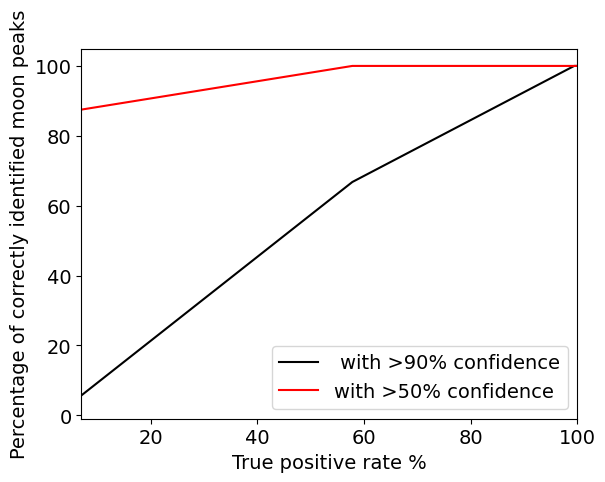}
    \caption{Percentage of correctly identified moon peaks with $>90\%$ confidence (black line) and $>50\% $ confidence (red line) as a function of the true positive rates, for Earth-sized moons with an orbital period of 26.4 hours.}
    \label{fig:truepositivevsconfidence}
\end{figure}

This section refers to the FAP calculations that are presented in Section \ref{Detectability} and shown in Figure \ref{fig:perioddogram}.
We calculated the FAPs and confidence intervals using a Gamma distribution with a scale parameter equal to the variance of the $95th$ percentile of the time-domain data and a shape parameter equal to one, as the square magnitude of each frequency coefficient for time domain data with Gaussian noise approximates a Gamma distribution.
Figure \ref{fig:truepositivevsconfidence} shows the percentage of correctly identified moon peaks that are detected with FAPs smaller than 10$\%$ and 50 $\%$ as a function of the true positive rate.

\end{appendix}

\end{document}